\documentclass[aps,prl,twocolumn,superscriptaddress]{revtex4-1}
\usepackage{amsmath}
\usepackage{amssymb}
\usepackage{graphicx}
\usepackage{bm}

\begin{document}

\title{Topological insulator in junction with ferromagnets: quantum Hall effects}

\author{A. L. Chudnovskiy}
\affiliation{1. Institut f\"ur Theoretische Physik, Universit\"at Hamburg,
Jungiusstr 9, D-20355 Hamburg, Germany}

\author{V. Kagalovsky}
\affiliation{ Shamoon College of Engineering, 
Basel/Bialik Sts, Beer-Sheva 84100, 
Israel }

\date{\today}

\begin{abstract}
The ferromagnet-topological insulator-ferromagnet (FM-TI-FM) junction exhibits thermal and electrical quantum Hall effects. The generated Hall voltage and transverse temperature gradient can be controlled by the directions of magnetizations in the FM leads, which inspires the use of FM-TI-FM junctions as electrical and as heat switches in spintronic devices. Thermal and electrical Hall coefficients are calculated as functions of the magnetization directions in ferromagnets and the spin-relaxation time in TI. Both the Hall voltage and the transverse temperature gradient decrease but are not completely suppressed even at very short spin-relaxation times. The Hall coefficients turn out to be independent of the spin-relaxation time for symmetric configuration of FM leads.  
\end{abstract}


\maketitle


The discovery and experimental realization of topological insulators (TI) opened a new and vividly developing field of theoretical and experimental  investigations \cite{Kane-Mele,Bernewig-Zhang,Koenig07,Hasan-Kane10}. Two-dimensional TI belong to the class of quantum spin Hall systems \cite{Kane-Mele,Bernewig-Zhang} that are distinguished by the existence of chiral spin-polarized edge states. There are two chiral states with opposite spin-projections at each edge  that propagate in the opposite directions. The existence of the edge states implies strong similarity between the properties of TI and of a quantum Hall system, although no external magnetic field is applied. Each spin-polarized edge state is subject to an effective magnetic field corresponding to a one magnetic flux quantum per electron (the condition for the lowest quantum Hall plateau), hence it contributes to the quantized Hall conductance of the sample. However, the signs of the effective magnetic fields are opposite for the edge states with opposite spin projections, which results in the exact cancellation of contributions from the two counter-propagating edge states to the total Hall conductance \cite{Bernewig-Zhang}.  

\begin{figure}
\includegraphics[width=8cm]{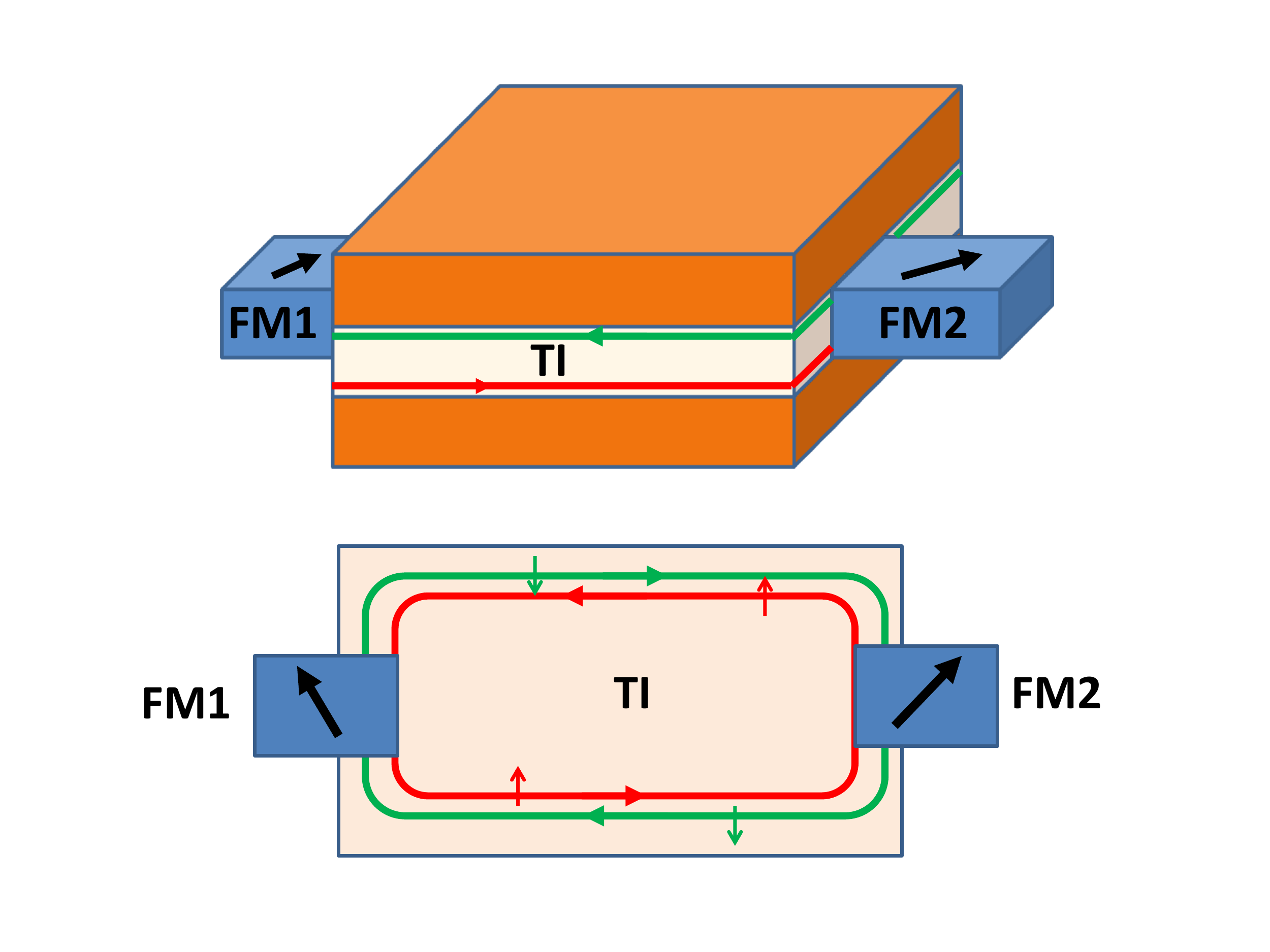} %
\vskip -0.1cm
\caption{(Color online) Proposed experimental setup of FM-TI-FM junction.  Spin-$\uparrow$ and spin-$\downarrow$ electrons have opposite chirality at the edge states. 
\label{fig-Setup}}
\end{figure}

It has been suggested in the early papers on quantum spin Hall effect that the properties of spin-polarized edge states can be probed by injecting spin-polarized currents in TI \cite{Bernewig-Zhang,Kane-Mele}.  
In this paper we show, that the quantum Hall electrical and thermal resistances can be revealed in the experimental measurements on  a two-dimensional TI sandwiched between the two ferromagnets (FM) in a FM-TI-FM junction (see Fig. \ref{fig-Setup}). The use of ferromagnets allows spin-selective contacting of the edge states in TI. In the ideal situation of completely polarized ferromagnets it is possible to contact a single chiral spin-polarized edge state. Thus one would measure quantized value of electrical Hall conductance $G_{\mathrm{Q}}=dI_{\parallel}/d V_H=e^2/h$ proper to the lowest Landau level of the integer quantum Hall effect. Similarly, the longitudinal heat flow through TI will result in the appearance of a transverse temperature gradient, which is the essence of the {\em thermal} Hall effect. The corresponding thermal Hall coefficient is also quantized $K_Q=dQ_{\parallel}/d T_{\perp}=(\pi^2 k_B^2/3 h)T$. 

\begin{figure}
\includegraphics[width=8cm]{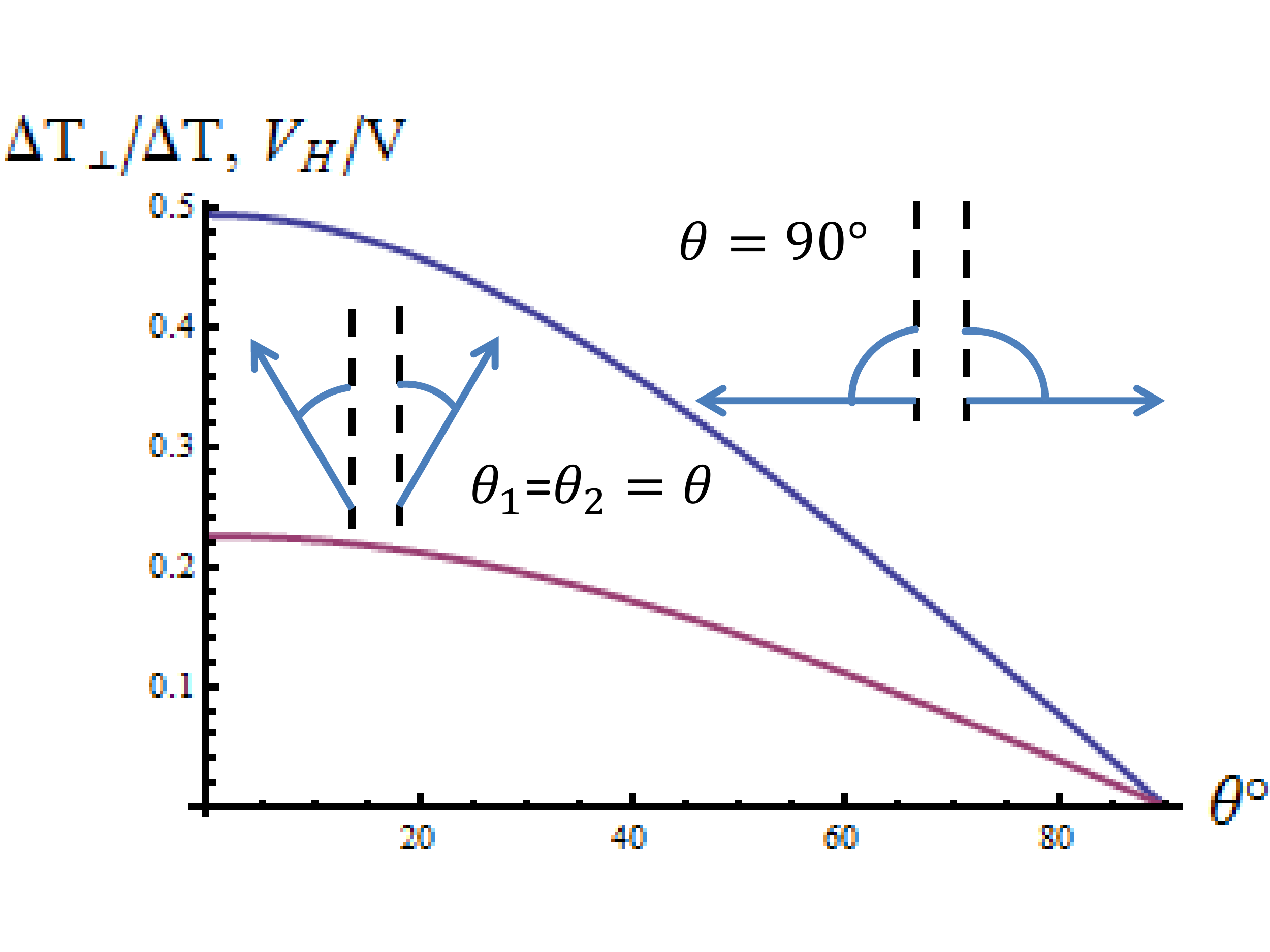}%
\vskip -0.4cm
\caption{(Color online) Ratio of the transverse to longitudinal  temperature gradient $\Delta T_{\perp}/ \Delta T$, which is equal to the ratio of  the Hall voltage to longitudinal voltage $V_{H}/V$, as a function of the angle $\theta$ between the magnetizations of ferromagnets and the spin-quantization axis in TI for  symmetric contacts. The polarizations of contacts are $p=1$ (upper curve) and $p=0.5$ (lower curve).  The spin-scattering time $\tau =10$. The total dimensionless conductances of the contacts  $g_{1}=g_{2}=1$.  Insets show the magnetization directions in ferromagnets (arrows) and the direction of the spin quantization axis in TI (dashed line) for the general symmetric configuration and the case of perpendicular orientation, equivalent to the absence of FM leads.  
\label{fig-eff_temperature_theta}}
\end{figure}

The coupling between the spin-polarized edge state of TI and FM lead depends on the angle between the magnetization of the lead and the direction of spin quantization axis in TI. The latter is determined by the crystallographic structure of TI \cite{Kane-Mele,Bernewig-Zhang,Koenig07,Hasan-Kane10}. Rotating the magnetization direction in FM leads, one can control the transverse voltage and the transverse thermal gradient induced in TI. It varies from a finite maximal value, when the orientation of magnetizations in FM is parallel to the spin-quantization axis in TI, to the complete suppression of transverse voltage and temperature gradient for the perpendicular orientation (see Fig. \ref{fig-eff_temperature_theta}). 

\begin{figure}
\includegraphics[width=8cm]{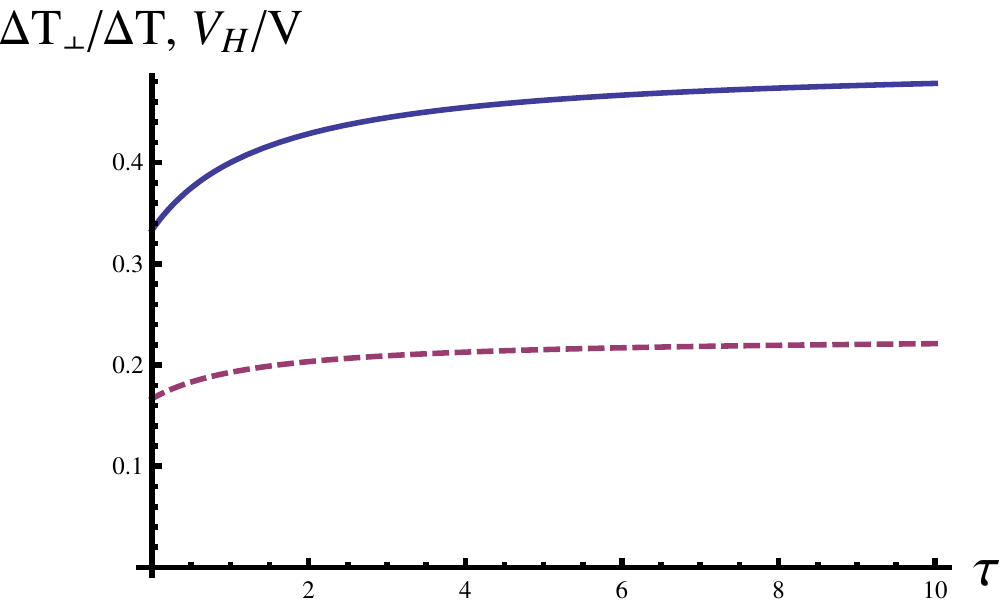} \\
\caption{Ratio of the temperature gradients and voltages $\Delta T_{\perp}/\Delta T=V_H/V$ as a function of the relaxation time $\tau$ for the parallel orientation of magnetizations $\theta=0$ and for the equal polarizations of ferromagnets $p=1$ (solid line) and $p=0.5$ (dashed line). The total dimensionless conductances of the contacts  $g_{1}=g_{2}=1$. 
\label{taug}}
\end{figure}

Topological insulators are often contaminated with magnetic impurities that introduce scattering between the chiral states at the edge.  
Nevertheless, as long as the localization length is larger than the system length \cite{Alt_Yud}, the edge states remain intact. 
The quasi-elastic spin-flip back-scattering by magnetic impurities, while reducing the transverse temperature gradient and the Hall voltage in general,  {\em does not suppress} the thermal and electrical Hall effects in FM-TI-FM structure completely (see Fig. \ref{taug}). The Hall coefficients remain finite even in the formal limit of infinitely short scattering time. 
The dimensionless electrical ($R_H=dV_H/dI_{\parallel}$) and thermal ($R_T= d T_{\perp}/dQ_{\parallel}$) Hall resistances are equal to each other,  
\begin{equation}
R_HG_Q=R_TK_Q=\mathcal{F}, 
\label{F}
\end{equation}
where the factor $\mathcal{F}$ depends on conductances, polarizations of ferromagnetic contacts, magnetization directions, and the spin-scattering time. Remarkably, in the case of identical ferromagnets with equal angle $\theta$ between the magnetization and spin-quantization axis of TI, the factor $\mathcal{F}$ turns out to be independent of the scattering time, its analytical expression reads,  
\begin{equation}
 \mathcal{F}(g,  p, \theta)=\frac{2p\cos\theta}{4-g+gp^2\cos^2\theta}, 
\label{Fsym}
\end{equation}
where $g$ denotes the total dimensionless conductance of each contact, and $p$ denotes the polarization of FM (to be defined below). 
For completely polarized ferromagnets ($p=1$) with  magnetizations parallel to the spin-quantization axis in TI ($\theta=0$), the Hall coefficients retain their quantized values.  At the same time, the ratio of transverse and  longitudinal voltages ($V_H/V$)  as well as transverse and longitudinal temperature gradients ($\Delta T_{\perp}/\Delta T$) depends on the scattering time, 
\begin{equation}
 \frac{V_H}{V}=\frac{\Delta T_{\perp}}{\Delta T}=\frac{t_0 n_0 v_0+1}{3+2t_0 n_0 v_0}, 
\label{ideal_case}
\end{equation}
where $t_0$ denotes the quasi-elastic  spin-flip scattering time by magnetic impurities, $n_0$ is the linear concentration of electrons on the edge, and  $v_0$ denotes the Fermi velocity in the edge state.

In the opposite case when electrons injected in the TI are completely unpolarized (FM polarization ($p=0$) or, equivalently, the magnetization of FM electrodes is perpendicular to 
the spin-quantization axis in the TI ($\theta=\pi/2$)) charge and thermal quantum Hall effects disappear in complete agreement with the situation in quantum spin Hall system \cite{Kane-Mele,Bernewig-Zhang}. In that case, the factor $\mathcal{F}$ in Eq. (\ref{Fsym}) equals to zero, indicating the vanishing Hall voltage and transverse temperature gradient.

\begin{figure}
\hskip -.5cm
\includegraphics[height=4cm]{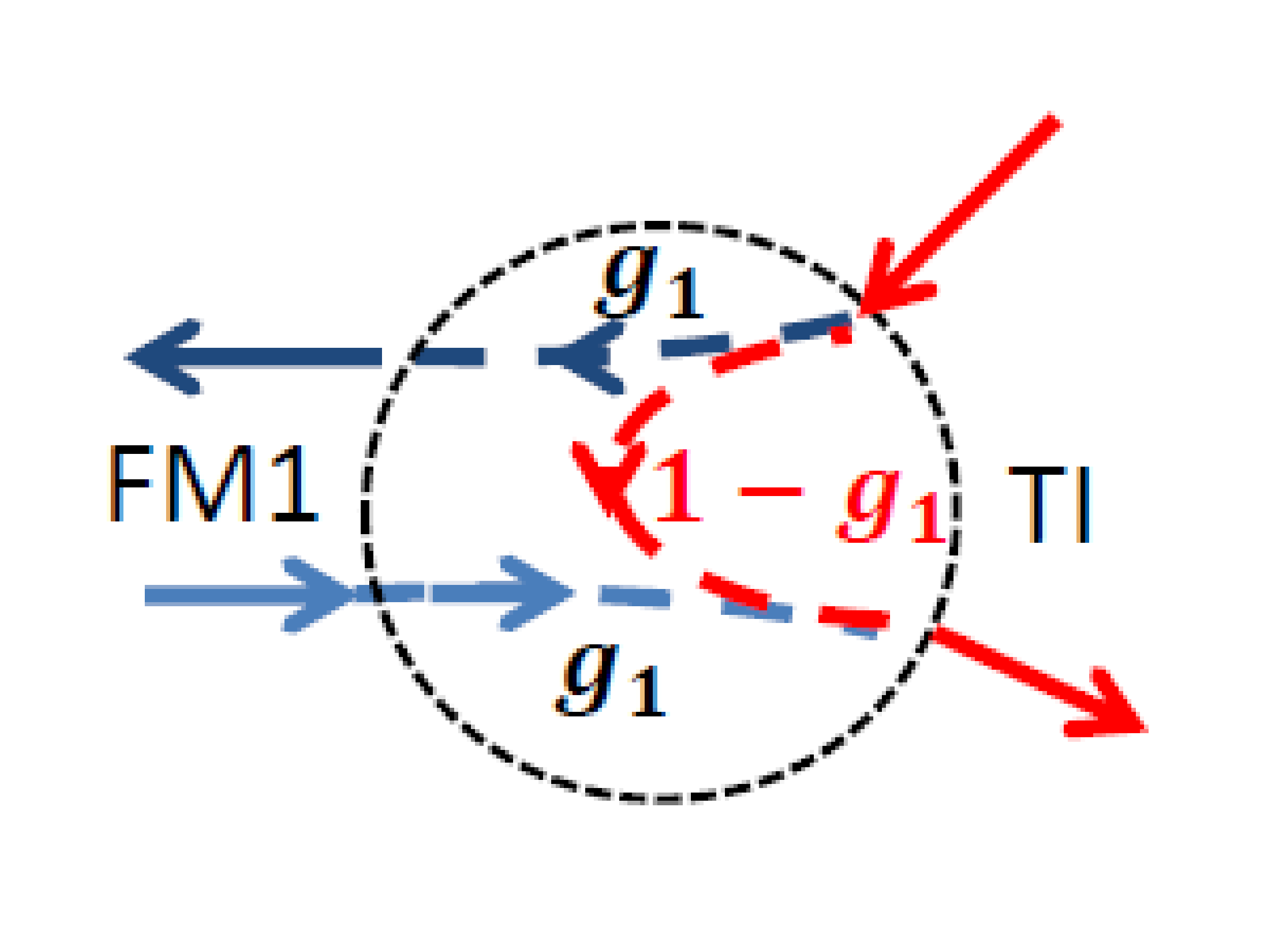}%
\vskip -.5cm
\caption{(Color online) Scheme of scattering in FM-TI contact for a single chiral spin-polarized edge state.    
\label{fig-Scatt}}
\end{figure}

The emergence of the temperature gradient transverse to the heat flow through TI is in fact identical to the thermal quantum Hall effect (Leduc-Righi effect) \cite{LL-Kinetics}.  Consider a single chiral edge state in TI, which we denote as the spin-up state:  Let us suppose that the left contact has a higher temperature than the right one ($T_1>T_2$) (see Fig. \ref{fig-Scatt}). In that case  the hot electrons from the left contact propagate along  the lower edge, and the cold electrons from the right contact propagate  along the upper edge. In the absence of relaxation, the electrons on the edges are not in the equilibrium, however, as it will be shown below,  one still can associate an effective temperature to the electron distribution. Thus,  a temperature difference between the edges is created that is perpendicular to the heat flow. At the same time, there is a counter-propagating spin-down edge state  in TI. For that state the Leduc-Righi effect has the opposite sign. If the reservoirs are spin-unpolarized, the temperature differences created by the spin-up and spin-down edge states compensate each other exactly resulting in zero net effect. Another situation is realized, if the reservoirs  are ferromagnetic. In that case the contact conductances for spin-up and spin-down electrons differ, the compensation of contributions from spin-up and spin-down edge states does not take place any more, and a finite temperature difference between the edges is predicted. Analogously, the Hall voltage is generated by the potential difference between the ferromagnets. 

In what follows we develop a general description of FM-TI-FM junction in terms of rate equations for distribution functions of the edge states. To this end let us consider the experimental setup shown schematically  in Fig. \ref{fig-Setup}. The contacts between ferromagnets and TI are described using Landauer-B\"uttiker scattering matrix formalism \cite{Buettiker}. 

Let us first assume no spin scattering between the edge channels of TI (no magnetic impurities). Then the spin channels do not mix and there is no inter-channel relaxation of the distribution functions, hence one can study the distribution of electrons in each spin-channel separately.  
Consider a single spin channel in more detail (see Fig. \ref{fig-Scatt}).  We suppress the spin index  for brevity.  
When the quantum coherence is preserved withing the TI edge state, then the propagation of electrons between the external reservoirs though TI should be considered as a quantum mechanical scattering problem. Introduce the tunnel and reflection amplitudes for each contact: $t_1, r_1, t_2, r_2$, $|t_i|^2+|r_i|^2=1$. Denote the annihilation operators in reservoirs as $a_1, a_2$, and in the upper and lower edge states as $a_{\mathrm{u}}, a_{\mathrm{d}}$.
 The scattering description of the contacts provides the following relations between the operators in reservoirs and in the edge channels 
\begin{eqnarray}
\nonumber
 a_{\mathrm{d}} e^{-i\phi/2}=r_1 a_{\mathrm{u}} e^{i\phi/2}+t_1 a_1 \\
 \nonumber
a_{\mathrm{u}}e^{-i\phi/2}=r_2 a_{\mathrm{d}}e^{i\phi/2}+t_2 a_2, \\
\end{eqnarray}
where $\phi$ is a phase collected by an electron along the edge, and solving for $a_{\mathrm{u}}$ and $a_{\mathrm{d}}$ we obtain 
\begin{eqnarray}
\nonumber
a_{\mathrm{u}}=\frac{t_1r_2}{1-r_1r_2e^{2i\phi}} a_1 e^{i\frac{3}{2}\phi}+ \frac{t_2}{1-r_1r_2e^{2i\phi}} a_2e^{i\frac{1}{2}\phi}, \\
\nonumber
a_{\mathrm{d}}=\frac{t_2r_1}{1-r_1r_2e^{2i\phi}} a_2 e^{i\frac{3}{2}\phi} + \frac{t_1}{1-r_1r_2e^{2i\phi}} a_1 e^{i\frac{1}{2}\phi}. \\
\label{a_up-down} 
\end{eqnarray}
Eq. (\ref{a_up-down}) immediately leads to the relation between the distribution functions on the edge and in the contacts. 
We denote the Fermi-Dirac distribution functions in the ferromagnets 
\begin{equation}
 F_i=\langle a_i^+ a_i\rangle, \ i=1,2.  
\end{equation}
The reservoirs are mutually uncorrelated, which implies $\langle a_1^+ a_2\rangle=\langle a_2^+ a_1\rangle=0$. Then the distribution functions in the upper anl lower edge channels can be expressed respectively as 
\begin{eqnarray}
&& \nonumber
 f_{\mathrm{u}}=\langle a_{\mathrm{u}}^+ a_{\mathrm{u}}\rangle=\frac{|t_1r_2|^2F_1+|t_2|^2F_2}{1+|r_1r_2|^2-2r_1r_2\cos(2\phi)}  , \\
&& \nonumber
f_{\mathrm{d}}=\langle a_{\mathrm{d}}^+ a_{\mathrm{d}}\rangle=\frac{|t_2r_1|^2F_2+|t_1|^2F_1}{1+|r_1r_2|^2-2r_1r_2\cos(2\phi)} . \\
\label{f_up-down}
\end{eqnarray}
The distribution functions Eqs. (\ref{f_up-down}) can also be expressed through dimensionless conductances of the contacts 
$g_i=(h/e^2)G_i$. Assuming that the phase $\phi$ is random (due to relaxation and thermalization processes) 
uniformly distributed variable, we average Eqs. (\ref{f_up-down}). Furthermore, taking into account that $|t_i|^2=g_i$, $|r_i|^2=1-g_i$ we finally get 
\begin{eqnarray}
\nonumber
f_{\mathrm{u}}=\frac{g_1(1-g_2)}{1+(1-g_1)(1-g_2)}F_1 + \frac{g_2}{1+(1-g_1)(1-g_2)} F_2,  \\
\nonumber
f_{\mathrm{d}}=\frac{g_2(1-g_1)}{1+(1-g_1)(1-g_2)}F_2 + \frac{g_1}{1+(1-g_1)(1-g_2)} F_1. \\
\label{f_up-down1}
\end{eqnarray}

As we have shown, when the phase coherence is lost on the length which is much shorter than the length of the edge channel, the contacts between the edge state and external reservoirs should be described in terms of transmission probabilities. We will now rederive our result in Eq. ({\ref{f_up-down1}}) and later use this approach to study the generic case of two spin-polarized edge channels. Consider, for example, an electron  coming from the upper edge to the contact 1 (see Fig. \ref{fig-Scatt}): 
it is absorbed into the lead FM$1$ with the probability $g_1$, which equals to the dimensionless contact conductance, and it is reflected from the contact into the lower edge with the probability $1-g_1$. At the same time, the incoming wave from  FM$1$ goes to the lower edge with the probability $g_1$. Analogous relations determine the scattering at the contact 2.   The time evolution of distribution functions in the upper and lower parts of the edge channel is governed by the following 
equations \cite{Buettiker} 
\begin{eqnarray}
&& \nonumber 
 \frac{df_u}{dt}=\frac{n_0 v_{0}}{2\pi}\left\{(1-g_2)f_d+g_2F_2-g_1f_u-(1-g_1)f_u\right\}, \\
&& \nonumber
\frac{df_d}{dt}=\frac{n_0 v_{0}}{2\pi}\left\{(1-g_1)f_u+g_1F_1-g_1f_d-(1-g_1)f_d\right\}. \\
\label{single-channel-eqs}
\end{eqnarray}
Since only elastic scattering is taken into account, the energy $\epsilon$ is conserved.  In writing Eqs. (\ref{single-channel-eqs}) we neglected the size of the region close to the contact, where the spatial change of the distribution function occurs, thus we omitted the spatial dependence of the distribution functions. 
The stationary solutions of Eqs. (\ref{single-channel-eqs}), $df_u/dt=0,  df_d/dt=0$ immediately reproduce Eqs. ({\ref{f_up-down1}}).  

The temperature difference or the finite voltage difference between the left ($F_1$) and right ($F_2$) reservoirs results in  non-equilibrium  distribution functions on the upper and the lower edge. Still one can define an effective temperature and an effective chemical potential at each edge by making a gedanken experiment, which consists of coupling each edge to a system in thermodynamical equilibrium, which we refer to as the thermometer. The temperature and the chemical potential of the thermometer, at which there is no net heat and particle flow between the thermometer and the edge can be defined as the effective temperature and the effective chemical potential \cite{Sivan-Imry,Pierre,Nazarov-Blanter}.   Therefore, the effective chemical potential and  the effective temperature on the edge $\nu=u,d$ are defined by the following equations,  
\begin{eqnarray}
 \int f_{\nu}(\epsilon) d\epsilon= \int \left[\exp\left(\frac{\epsilon-\mu_{\nu}}{k_B T_{\nu}}\right)+1\right]^{-1} d\epsilon   , \label{mu_eff} \\
\int \epsilon f_{\nu}(\epsilon) d\epsilon= \int\left[\exp\left(\frac{\epsilon-\mu_{\nu}}{k_B T_{\nu}}\right)+1\right]^{-1} \epsilon d\epsilon, \label{T_eff} 
\end{eqnarray}
where we assumed that the density of states in the conductance channel does not depend on energy. Experimentally, the measurement of a single spin channel can be realized if both the leads and the thermometers are completely polarized ferromagnets. 

Applying Eqs. (\ref{mu_eff}) and (\ref{T_eff}) to the distribution functions Eqs. (\ref{f_up-down1}), one obtains the relations between the longitudinal ($\Delta T=T_1-T_2$) and transverse temperature ($\Delta T_{\perp}$) differences as well as between the longitudinal and Hall voltages in the form,  
\begin{equation}
\frac{\Delta T_{\perp}}{\Delta T}=\frac{V_H}{V} =\frac{g_1g_2}{1-(1-g_1)(1-g_2)}. 
\label{HallRatios}
\end{equation}
In this case we obtain the electrical and thermal Hall conductances proper to the first integer quantum Hall plateau  $G_{\mathrm{H}}=G_Q$,  $K=K_Q$. 

Now let us turn to the generic case of two spin-polarized edge channels. In that case, the contact between a FM lead and each spin-polarized channel is characterized by a spin-dependent dimensionless conductance $g_{i\sigma}$, where $i=1,2$ corresponds to the FM lead and $\sigma=\uparrow,  \downarrow$ denotes the spin-projection of the edge state. The total dimensionless conductance of the contact is given by the sum $g_{i}=g_{i\uparrow}+g_{i\downarrow}$. 
 The angular dependence of contact conductances stems from the tunneling magnetoresistance effect \cite{Slonczewski-Sun2007}.  The band structure of a ferromagnet consists of majority (further denoted as ``+'') and minority (further denoted as ``-'') spin-polarized bands, that have different density of states (DoS) at the Fermi level. Therefore, each spin-polarized band provides a different contribution to the contact conductance.  Analogously, there are spin-polarized energy bands for the edge-states in TI, although their DoS are equal. If the magnetization direction in FM is parallel to the spin-quantization axis in TI, each spin-polarized band of FM couples to a single spin-polarized edge state of TI.  For an arbitrary angle $\theta$ between the magnetization in the  ferromagnet and the spin-quantization axis in the TI, each band in FM has finite hybridizations with both edge states of TI. Thereby the hybridization strength depends on $\theta$. This in turn results in the angular dependence of partial 
conductances between the FM band ($+/-$) and the spin-polarized edge state ($\uparrow/\downarrow$), which is given by the expressions \cite{Slonczewski-Sun2007}
\begin{eqnarray} 
\nonumber && 
G_{\uparrow}^{ +}(\theta)=G^{+}\cos^2(\theta/2), \ G_{\downarrow}^{ +}(\theta)=G^{+}\sin^2(\theta/2) \\
 && 
G_{\uparrow}^{ -}(\theta)=G^{-}\sin^2(\theta/2), \ 
G_{\downarrow}^{ -}(\theta)=G^{-}\cos^2(\theta/2). 
\label{G_theta}
\end{eqnarray}
Since both edge states of TI have the same DoS, the partial conductances Eqs. (\ref{G_theta}) are determined by only two independent material parameters,  $G^{+}$ and $G^{-}$, that characterize the coupling to the majority and minority bands respectively. 
It is convenient to characterize the spin-selectivity of the contact ($i$) by the contact polarization degree of the ferromagnet $p_i$, that is defined as  
\begin{equation}
 p_i=\sum_{\sigma=\uparrow, \downarrow}\left(G_{i\sigma}^{+}-G_{i\sigma}^{-}\right)/G_i, 
\label{p}
\end{equation}
where $G_i=\sum_{\sigma=\uparrow, \downarrow}(G_{i\sigma}^{+}+G_{i\sigma}^{-})$ is the total conductance of the contact. Thus, the total angular-dependent conductances to the spin-up and spin-down edge channels are given by 
\begin{equation}
G_{i\sigma}= G_i(1\pm p_i \cos\theta_i)/2, 
\label{Gsigma}
\end{equation}
where the upper and the lower sign corresponds to spin-up and spin-down edge state respectively. 
Analogous expressions follow for the dimensionless partial conductances $g_{i\sigma}=(h/e^2)G_{i\sigma}$. 

Therefore, a FM-TI-FM junction provides a possibility to create different contact conductances for different spin-channels and thus get a controlled net thermal and electrical Hall conductances of the whole system. Moreover, edge states can be contacted individually in the case of completely polarized ferromagnet that can be realized in semi-magnetic semiconductors. In that case the electrical and thermal Hall conductances proper to the integer quantum Hall effect are observed.  

Electrons in each edge state are described by a distribution function $f_{\nu\sigma}$, where $\nu$ denotes the position of the  edge ($\nu=u, d$), and $\sigma=\uparrow, \downarrow$ is a spin-index, which also determines the chirality of the edge state  in TI.   Spin-scattering  between the counter-propagating edge states induced by  magnetic impurities is taken into account phenomenologically, introducing the spin-scattering time $t_0$ \cite{Tanaka2011}. 
Thus, the rate equation describing the evolution of the distribution function at the edge $\nu$ with spin $\sigma$ can be written as 
\begin{equation}
\frac{d f_{\nu\sigma}}{dt}=\frac{n_0 v_{0}}{2\pi}\left\{(1-g_{\nu\sigma})f_{\bar{\nu}\sigma}-f_{\nu\sigma}+\Gamma_{\nu\sigma}\right\}
-\frac{1}{t_0}(f_{\nu\sigma}-f_{\nu\bar{\sigma}}), 
\label{RateEq}
\end{equation}
 where  $\bar{\nu}$ denotes the edge opposite to $\nu$ and $\bar{\sigma}$ denotes the spin projection opposite to $\sigma$.  According to  the chirality of the edge states (cf. Figs. \ref{fig-Setup}, \ref{fig-Scatt}), we define 
\begin{eqnarray}
&&
 g_{u\uparrow}=g_{2\uparrow}, \ g_{u\downarrow}=g_{1\downarrow}, 
 g_{d\uparrow}=g_{1\uparrow}, \ g_{d\downarrow}=g_{2\downarrow}, \\ 
&& \Gamma_{u\uparrow}(\epsilon)=g_{2\uparrow}F_{2}(\epsilon),  \ 
 \Gamma_{u\downarrow}(\epsilon)=g_{1\downarrow}F_{1}(\epsilon), \\
&&\Gamma_{d \uparrow}(\epsilon)=g_{1\uparrow}F_{1}(\epsilon),  \  
\Gamma_{d\downarrow}(\epsilon)=g_{2\downarrow}F_{2}(\epsilon). 
\label{Gamma_R}
\end{eqnarray}

Stationary solutions of Eq. (\ref{RateEq}) contain the full information about the nonequilibrium distribution functions of the edge states. The analytical solution can be represented in the form
\begin{equation}
 f_{\mu\sigma}(\epsilon)=\mathcal{D}_{\mu\sigma}(\epsilon)/D_0, 
\label{sol_gen}
\end{equation}
where 
\begin{eqnarray}
\nonumber && 
 D_0=(g_{1\uparrow}+g_{2\uparrow}-g_{1\uparrow}g_{2\uparrow})(g_{1\downarrow}+g_{2\downarrow}-g_{1\downarrow}g_{2\downarrow})\\ 
\nonumber && 
+\left(g_0-g_{1\uparrow}g_{2\uparrow} -g_{1\downarrow}g_{2\downarrow}\right)/\tau \\ 
&& 
+\left[2g_0-(g_{1\uparrow}+g_{1\downarrow})(g_{2\uparrow}+g_{2\downarrow})\right]/\tau^2, 
\label{D0}
\end{eqnarray}
and 
\begin{eqnarray}
\nonumber &&
 \mathcal{D}_{\mu\sigma}(\epsilon)=\left[\Gamma_{\mu\sigma}(\epsilon)+(1-g_{\mu\sigma})\Gamma_{\bar{\mu}\sigma}(\epsilon)\right] (g_{\mu\bar{\sigma}}+g_{\bar{\mu}\bar{\sigma}}-g_{\mu\bar{\sigma}}g_{\bar{\mu}\bar{\sigma}}) \\
\nonumber && 
+\left[2\Gamma_0(\epsilon)+\Gamma_{\mu\sigma}(\epsilon) (g_{\mu\bar{\sigma}}+g_{\bar{\mu}\bar{\sigma}} -g_{\mu\bar{\sigma}}g_{\bar{\mu}\bar{\sigma}})-2\Gamma_{\bar{\mu}\sigma}(\epsilon) g_{\mu\sigma} \right. \\
\nonumber && 
\left. 
 -\Gamma_{\mu\bar{\sigma}}(\epsilon) (g_{\mu\sigma}+g_{\bar{\mu}\bar{\sigma}}-g_{\mu\sigma}g_{\bar{\mu}\bar{\sigma}})
 -\Gamma_{\bar{\mu}\bar{\sigma}}(\epsilon) (g_{\mu\sigma}+g_{\mu\bar{\sigma}})\right]/\tau\\
&& 
+\left[2\Gamma_0(\epsilon)-(\Gamma_{\bar{\mu}\sigma}(\epsilon)+\Gamma_{\bar{\mu}\bar{\sigma}}(\epsilon))(g_{\mu\sigma}+g_{\mu\bar{\sigma}})\right]/\tau^2
\label{D_mu-sigma}
\end{eqnarray}
with $g_0=\sum_{\nu\sigma}g_{\nu\sigma}$,  $\Gamma_0=\sum_{\nu\sigma}\Gamma_{\nu\sigma}$.   Here we introduced the dimensionless relaxation time $\tau=t_0 n_0 v_0$.

To calculate the thermal Hall coefficient and the Hall conductance, we need expressions for the heat flow and the electrical current through the system. In the stationary state, the heat flow and the electrical current can be related to the particle flow at one of the contacts. For example, the particle flow out of the contact 1 (the left contact in Figs. \ref{fig-Setup}, \ref{fig-Scatt}) consists of spin-up electrons going from the contact to the lower edge, and spin-down electrons going  from the contact to the upper edge. Those particles have the distribution function of FM1. The flow into the contact consists of spin-up electrons coming from the upper edge and  spin-down electrons coming  from the lower edge, which are distributed according to the distribution functions $f_{u\uparrow}$ and $f_{d\downarrow}$ respectively. Therefore, the total electrical current and the heat flow through the contact are given by 
\begin{eqnarray*}
&& 
 I=\frac{e}{h}\int d\epsilon \left\{\left(g_{1\uparrow} + g_{1\downarrow}\right) F_1(\epsilon)-g_{1\uparrow} f_{u\uparrow}(\epsilon) -g_{1\downarrow}f_{d\downarrow}(\epsilon)\right\}, \label{I}\\ 
&&
 Q=\int \epsilon d\epsilon \left\{\left(g_{1\uparrow} + g_{1\downarrow}\right) F_1(\epsilon)-g_{1\uparrow} f_{u\uparrow}(\epsilon) -g_{1\downarrow}f_{d\downarrow}(\epsilon)\right\}. \label{Q}
\end{eqnarray*} 

To calculate the Hall voltage and the transverse temperature gradient for the case of two spin edge channels we define effective chemical potential and temperature at each edge $\nu$ (analogously to Eqs. (\ref{mu_eff},
\ref{T_eff})

\begin{eqnarray*}
&& 
 \int (f_{\nu\uparrow}(\epsilon)+f_{\nu\downarrow}(\epsilon)) d\epsilon= \int \left[\exp\left(\frac{\epsilon-\mu_{\nu}}{k_B T_{\nu}}\right)+1\right]^{-1} d\epsilon   , \label{I1}\\ 
&&\int (f_{\nu\uparrow}(\epsilon)+f_{\nu\downarrow}(\epsilon)) \epsilon d\epsilon= \int \left[\exp\left(\frac{\epsilon-\mu_{\nu}}{k_B T_{\nu}}\right)+1\right]^{-1} d\epsilon   \label{Q1}\\ 
\end{eqnarray*}

Calculating the relations between the transverse temperature gradient and the longitudinal heat current $R_T=\Delta T_{\perp}/Q$ and also between the Hall voltage and  the electrical current $R_{H}=V_{H}/I$, we obtain the dependence of the Hall coefficients on the parameters of the experimental setup. Thereby the dimensionless thermal and electrical Hall coefficients turn out to be equal to each other, as stated in Eq. (\ref{F}). Moreover, the Hall coefficients become independent of the relaxation time $\tau$ for the symmetric contacts, as it is pointed out in Eq. (\ref{Fsym}). 

Eqs. (\ref{sol_gen}) -- (\ref{D_mu-sigma}) give the distribution functions of the spin-polarized edge channels as linear combinations of the distribution functions $F_1(\epsilon)$, $F_2(\epsilon)$ in ferromagnets with coefficients depending on partial conductances  of the contacts $g_{\nu\sigma}$ and the relaxation time $\tau$. They  simplify substantially  in particular limiting cases. Consider the case of completely polarized ferromagnets, $p_1=p_2=1$, with perfect contact conductances $g_1=g_2=1$,  and equal angle $\theta$ between the magnetization of the left and the right ferromagnet, and the quantization axis in TI.  In that case  we obtain 
\begin{eqnarray}
\nonumber && 
f_{u \uparrow}(\epsilon)=\frac{(6+ 2\cos\theta)(1+\tau)}{12+9\tau-\tau\cos^2\theta}F_{2}(\epsilon) \\
&&
+\frac{[3(2+\tau)- 2(1+\tau)\cos\theta 
-\tau\cos^2\theta] }{12+9\tau-\tau\cos^2\theta} F_{1}(\epsilon). 
\label{Sol_G=1}
\end{eqnarray}
The distribution function $f_{u\downarrow}$ is obtained from Eq. (\ref{Sol_G=1}) by interchange $F_1(\epsilon)\leftrightarrow F_2(\epsilon)$, and $\cos\theta \rightarrow -\cos\theta$. 
Distribution functions on the lower edge $f_{d\sigma}$ are obtained from the distribution functions  $f_{u\sigma}$ by interchange $F_1(\epsilon)\leftrightarrow F_2(\epsilon)$.

In the absence of relaxation, $\tau\rightarrow\infty$, and for magnetizations of ferromagnets parallel to the spin-quantization axis in TI, $\theta=0$, the spin-up edge states acquire the distribution functions from the corresponding FM leads, 
$f_{u \uparrow}(\epsilon)=F_{2\uparrow}(\epsilon)$, $f_{d \uparrow}(\epsilon)=F_{1\uparrow}(\epsilon)$, whereas the electron distribution for the spin-down states is the same on both edges, and it equals the half-sum of the distribution functions in the leads, 
$f_{u \downarrow}(\epsilon)=f_{d \downarrow}(\epsilon)=\left(F_{1\uparrow}(\epsilon)+F_{2\uparrow}(\epsilon)\right)/2$. Experimental measurement of the transverse temperature gradient performed with completely spin-polarized probes would reveal the ideal result $\Delta T_{\perp}=\Delta T$, and hence it would give the quantum thermal Hall coefficient 
$K_Q$. 
Measurement of the temperature gradient with a spin-unpolarized thermometer will effectively average the electron distributions of the spin-up- and spin-down-states at each edge, leading to the result $\Delta T_{\perp}=\Delta T/2$, hence $R_T=1/(2K_Q)$. Analogous results follow for the Hall voltage. Measurements of the Hall voltage by spin-polarized probes would give the ideal result $V_H=V$, and the quantized Hall conductance $G_Q$, while the measurements with nonmagnetic probes would give $V_H=V/2$, $R_H=1/(2G_Q)$. Those results are in complete analogy to the measurements of the quantum Hall effect with spin-degenerate vs. spin-splitted Landau levels \cite{vonKlitzing}. 

In the opposite case of very fast relaxation (within the validity of our model as discussed above), $\tau\rightarrow 0$,  the spin-polarized states at each edge are strongly mixed, and their distribution functions become equal 
\begin{eqnarray}
\nonumber 
f_{u \uparrow}(\epsilon)=f_{u \downarrow}(\epsilon)=\frac{F_1(\epsilon)+F_2(\epsilon)}{2}
-\frac{F_1(\epsilon)-F_2(\epsilon)}{6} \cos\theta, \\
\nonumber
f_{d \uparrow}(\epsilon)=f_{d \downarrow}(\epsilon)=\frac{F_1(\epsilon)+F_2(\epsilon)}{2}
+\frac{F_1(\epsilon)-F_2(\epsilon)}{6}\cos\theta.
\end{eqnarray}
Thus, we obtain the following result for the transverse temperature gradient and for the Hall voltage,  
\begin{equation}
 \Delta T_{\perp}/\Delta T=V_H/V=(1/3)\cos\theta. 
\label{Hall_tau0}
\end{equation} 
Because of a strong mixing between the spin-up and spin-down states, the measurements with a spin-polarized and spin-unpolarized thermometer would give the same result. Interestingly, both the transverse temperature difference and the Hall voltage do not disappear  even for the fast spin relaxation, meaning that the thermal and electrical Hall effects are not suppressed by the quasi-elastic scattering by magnetic impurities.  Their dependence on the spin-relaxation time is shown in Fig. \ref{taug}.

In conclusion, we showed theoretically that the thermal quantum  Hall effect as well as the electrical quantum Hall effect can be observed in topological insulators in contact with ferromagnetic leads. By changing directions of the magnetizations in FM  with respect to the spin-quantization axis in TI, one can obtain a large degree of  control over the generated Hall voltage and transverse temperature gradient. 
The measured Hall effects are maximal, if the measurement is performed by ferromagnetic probes, reaching the quantized  values of electrical and thermal Hall coefficients for completely spin-polarized FM leads and thermometers. 
Measurements by magnetically unpolarized probes give smaller values of  the Hall voltage and temperature gradient but the effect remains substantial. 
Quasi-elastic back-scattering by magnetic impurities in TI reduces the induced Hall voltage and transverse temperature gradient, although it {\em does not} suppress the effect completely. The thermal and electrical Hall coefficients remain  finite 
even in the limit of very fast spin-relaxation time. The experimental setup proposed in this Letter lies well within the reach of modern technology  \cite{Deviatov}. Of special importance for experimental measurements is the symmetric configuration of FM leads, in which case the Hall coefficients are independent of disorder. This work gives impetus to the experimental realization of FM-TI-FM devices and their application in spintronics.   

\begin{acknowledgments}
Authors acknowledge support from DFG through the Priority Program 1285 ``Semiconductor Spintronics'', and from SCE internal grant.  
\end{acknowledgments}


\begin{thebibliography}{10}
\bibitem{Kane-Mele} C. L. Kane, E. J. Mele, Phys. Rev. Lett. {\bf 95}, 146802, (2005) C. L. Kane, E. J. Mele, ibid., {\bf 95}, 226801 (2005).
\bibitem{Bernewig-Zhang}  B. Andrei Bernevig and Shou-Cheng Zhang, Phys. Rev. Lett. {\bf 96}, 106802 (2006). 
\bibitem{Koenig07}M. K\"onig, S. Wiedmann, C. Brne, A. Roth, H. Buhmann, L.
W. Molenkamp, X. L. Qi and S. C. Zhang, Science {\bf 318}, 766 (2007).
\bibitem{Hasan-Kane10} M. Z. Hasan, C. L. Kane, Rev. Mod. Phys. {\bf 82}, 3045 (2010).
\bibitem{Alt_Yud} B. L. Altshuler, I. L. Aleiner, and V. I. Yudson, Phys. Rev. Lett.  {\bf 111}, 086401 (2013). 
\bibitem{LL-Kinetics} E.M. Lifshitz and L.P. Pitaevskii, {\it Physical Kinetics} (Pergamon Press, New York 1981), p. 244.
 \bibitem{Buettiker} M. Buttiker,  Y. Imry, R. Landauer, S. Pinhas, Phys. Rev. B {\bf 31}, 6207 (1985); 
 M. B\"uttiker, Phys. Rev. B {\bf 38}, 9375 (1988). 
\bibitem{Sivan-Imry}U. Sivan and Y. Imry, Phys. Rev. B {\bf 33}, 551 (1986). 
\bibitem{Pierre} C. Altimiras, H. le Sueur, U. Gennser, A. Cavanna, D. Mailly and F. Pierre, Nature Physics {\bf 6}, 34 (2010).
\bibitem{Nazarov-Blanter} Y. V. Nazarov, Y. M. Blanter, {\it Quantum Transport. Introduction to Nanoscience} (Cambridge University Press, New York 2009) pp. 54, 158. 
\bibitem{Slonczewski-Sun2007} J. C. Slonczewski and J. Z. Sun, J. Magn. Magn. Mater. {\bf 310}, 169 (2007). 
\bibitem{Tanaka2011} Y. Tanaka, A. Furusaki, and K. A. Matveev Phys. Rev. Lett. {\bf 106}, 236402 (2011).
\bibitem{vonKlitzing} S. Koch, R. J. Haug, K. v. Klitzing, and K. Ploog, Phys. Rev. Lett. {\bf 67}, 883, (1991). 
\bibitem{Deviatov} A. Kononov, S.V. Egorov, G. Biasiol, L. Sorba, and E.V. Deviatov, arXiv:1401.5719 (2014). 

\end{thebibliography}
\end{document}